%% file: main.tex
\documentclass[a4paper,french]{rnti}

\usepackage[T1]{fontenc}
\usepackage[frenchb]{babel}
\usepackage{graphicx}
\usepackage{caption}
\usepackage{subcaption}
\usepackage{multirow}

\usepackage{url}

\usepackage{graphicx}

\usepackage{listings}
\usepackage{latexsym}

\usepackage{fancyhdr}
  
\newcommand{\logs}{\textit{logs}}

\makeatletter
\def\input@path{{./tex/}}
\graphicspath{{./figures/}}

\titrecourt{Automatisation de la structuration des logs pour le Cloud Computing}

\nomcourt{I. Vervaet et al.}

\titre{Automatisation de la structuration des logs pour le Cloud Computing}

\auteur{Arthur Vervaet\affil{1},
        Raja Chiky\affil{2},
        Mar Callau-Zori\affil{1}}

\affiliation{
    \affil{1}3DS Outscale,1 rue Royale 92210 Saint-Cloud\\
    \affil{2}LISITE-ISEP, 10 rue de Vanves, 92130 Issy-les-Moulineaux\\
    arthur.vervaet@outscale.com, raja.chiky@isep.fr\\ 
 }

\resume{%
Les registres de \textit{logs} sont une composante fondamentale des systèmes informatiques modernes. Ils
permettent aux équipes d'analyse et de surveillance de comprendre les comportements anormaux ou malveillants ayant
pu survenir. Cependant, l'augmentation permanente du volume de \textit{logs} générés par ces systèmes a rendu impossible
l'inspection manuelle et pose un véritable défi d'automatisation du processus.
Afin de traiter automatiquement ces données, plusieurs solutions de structuration des \textit{logs} ont vu le jour.
Dans cet article, nous analysons les capacités de deux d'entre elles à répondre aux enjeux du \textit{Cloud Computing} en termes d'efficience et d'efficacité. 
Nos travaux se concentrent sur l'impact des paramètres et
du prétraitement sur les performances de ces méthodes, deux étapes importantes, car nécessitant une intervention humaine incompatible
avec l'automatisation du processus de structuration des \textit{logs}.
}

\summary{%
Logs are a fundamental component of modern computer systems.
They enable the analysis and monitoring teams to understand any abnormal or malicious
behavior that may have occurred. The continuous increase in the volume of logs generated by these systems made
it unsuitable for manual inspection and represents a real challenge with regard to process automation. In order to process
these data, several log-structuring solutions have been developed. 
In this article, we analyze the capabilities of two solutions in order to meet the challenges of Cloud Computing in terms of efficiency and effectiveness. 
Our work focuses on the impact of parameterization and preprocessing on the performance of these methods-- two important steps as 
they require human intervention, which is incompatible with with the automation of the log-structuring process.
}

\begin{document}

\input{intro}
\input{problem}
\input{setup}
\input{parameters}
\input{preprocessing}
\input{timeprocessing}
\input{conclusion}

\bibliographystyle{rnti}
\bibliography{biblio}

\appendix

\Fr

\end{document}

%% file: intro.tex
\section{Introduction}\label{sec:intro}

Les plateformes de \textit{Cloud Computing} mettent à disposition de leurs clients différentes ressources informatiques à la demande. Cette externalisation rend les fournisseurs garants
de la haute disponibilité et de la qualité de leurs services.
La gestion d'un parc de ressources mutualisées en croissance constante demande de minimiser l'intervention humaine afin de suivre le changement d'échelle des infrastructures et d'éviter les erreurs.
Pour atteindre cet objectif, on doit  pouvoir se servir de toutes les informations à disposition afin de développer des outils autonomes servant à contrôler et assurer le respect de la qualité de service.

La journalisation des événements ou \logs{}, consiste à enregistrer de manière détaillée des informations relatives à un programme pendant son exécution. Ces \logs{} constituent une source d’information précieuse, utilisée pour retracer les différentes étapes d’un processus à la recherche de l’origine d'erreurs ou de pannes, mais également pour identifier des anomalies de performance ou analyser les statistiques d'utilisation~\citep{zhu2019tools}.

Malgré toutes les possibilités offertes par les \logs{} et l'information qu'ils contiennent, les traiter efficacement est une tâche complexe. Les développeurs ayant très peu de contraintes dans l'écriture des macros destinées à produire les messages \logs{}, ceux-ci suivent un format semi-structuré.
Par exemple, le protocole RFC 5424 pour rsyslog~\citep{rfc5424} fixe un format en trois parties: un en-tête (HEADER), une partie optionnelle pour des données structurées (STRUCTURED-DATA) et un message (MSG).
À l'inverse du HEADER qui possède un format prédéfini contenant la priorité et l'origine du message, le champ MSG doit uniquement contenir un message libre fournissant des informations sur l'événement.

Du fait de la complexité croissante et du besoin de scalabilité des plateformes Cloud, la volumétrie des \logs{} ne fait qu'augmenter, et ce de façon rapide. En 2013, Alibaba produisait environ 30 à 50 gigaoctets (environ 120 à 200 millions de lignes) par heure~\citep{mi2013toward}. ~\citet{zhu2019tools} évoquent un système de Huawei produisant plusieurs téraoctets de \logs{} par jour en 2019.
De notre côté, nous produisons actuellement chez 3DS OUTSCALE (fournisseur de Cloud français) une moyenne de 250 000 \logs{} par seconde dans un de nos systèmes.
Le traitement de cette volumétrie demande un système autonome nécessitant le moins d'intervention humaine possible. Afin de réagir le plus rapidement possible à une potentielle dégradation du service fourni, ce système doit pouvoir fonctionner en temps quasi réel.

L'importance de la structuration des \logs{} a motivé la recherche et durant ces dernières années plusieurs comparatifs des algorithmes existants ont vu le jour~\citep{he2016evaluation, zhu2019tools}.
Ces études évaluent la précision et la robustesse de différentes solutions de structuration des \logs{} sur des jeux de données provenant d'applications différentes.
  Les résultats de ces études illustrent l'utilisation possible de méthodes existantes pour le traitement des \logs en temps réel. Cependant, à notre connaissance aucune étude ne pointe les limites restantes pour une automatisation complète du processus.
En utilisant deux méthodes de traitement de \logs, opérant en ligne et présentant de bons résultats, nous avons étudié deux points indispensables à leur automatisation: l'indépendance de ces méthodes au paramétrage et au prétraitement.

Nos contributions sont : une étude de l'impact du paramétrage sur la précision des méthodes et une étude de l'influence du prétraitement sur le temps de traitement et la précision.
Ces travaux illustrent la faible dépendance d'une des méthodes étudiées au paramétrage
avec cependant un impact modéré, mais potentiellement négatif du prétraitement sur sa précision. A contrario de la seconde méthode étudiée pour laquelle la précision et le temps
de traitement dépendent fortement des étapes étudiées.

Dans ce qui suit, nous présentons l'état de l'art pour la structuration de \logs (Sec.~\ref{sec:problem}), dans la Sec.~\ref{sec:relatedwork} nous présentons d'autres travaux comparatifs des algorithmes existants.
La Sec.~\ref{sec:evaluation} présente notre étude des caractéristiques fondamentales pour un système autonome dans le Cloud : la précision et le temps de traitement en fonction du paramétrage et du prétraitement.
Enfin, nous concluons le document dans la Sec.~\ref{sec:conclusion}.

%% file: problem.tex
\section{Structuration des \logs{} }\label{sec:problem}

La structuration des \logs{} se focalise sur la partie libre du message et cherche à identifier deux composantes :
1/ une composante fixe qui sert de patron
2/ une composante variable, contenant les spécificités du message.
Dans le message "{ \tt New process started: process x92 started on port 42}", le patron identifié par l'expression régulière est "{\tt New process started: process .* started on port .*}"; le contenu spécifique est représenté par la liste: {\tt [x92, 42]}.

Trois qualités sont primordiales dans la structuration des \logs{}: la précision, la robustesse et l'efficacité~\citep{zhu2019tools}.
La précision évalue la bonne détection des différentes parties d'un message.
Dans un système où la volumétrie des \logs{} à traiter est conséquente et peut évoluer rapidement, il est nécessaire de disposer de solutions robustes, à même de maintenir leurs performances dans un environnement changeant.
Ces solutions doivent également être efficaces et capables de traiter les messages en temps quasi réel.

Dans un cadre industriel, il est humainement coûteux et pas toujours possible d'obtenir un jeu de données labellisées pour paramétrer un système, d'où la nécessité de restreindre l'impact de cette étape sur les performances.
Il est courant d'utiliser la connaissance d'un système pour commencer à structurer ses \logs{} à l'aide d'expressions régulières servant à identifier des composantes variables.
À l'aide d'expressions simples, on peut identifier le chiffre et l'adresse IP du message "{\tt Send 92 bytes to system x32 at 112.13.92.1}"
et ainsi obtenir un message prétraité:  "{\tt Send * bytes to system x32 at *}", les * symbolisant l'emplacement d'une variable identifiée.
L'inconvénient de ce prétraitement est qu'il demande une connaissance précise des données traitées et l'intervention d'un ou plusieurs experts pour définir les expressions régulières à employer. De plus, le prétraitement peut impacter négativement la performance (Section~\ref{sec:evaluation}).

Les \logs{} étant générés sous forme de flux, le meilleur moyen de les traiter est de suivre le même fonctionnement. Plusieurs {\it log parsers} présentés dans la littérature permettent cela.
On retrouvera SHISO~\citep{mizutani2013incremental} qui se base sur un arbre de structuration enrichi durant le fonctionnement,
LenMa~\citep{shima2016length} qui utilise la séquence constituée par la longueur de chaque mot dans un \textit{log} pour classifier une entrée,
Spell~\citep{makanju2011lightweight} basé sur la recherche de la plus longue séquence commune pour rattacher un \textit{log} à un patron déjà connu ou en découvrir un nouveau, et enfin
Drain~\citep{he2017drain} qui construit un arbre de partitionnement de profondeur fixe pour structurer les \logs{}.

\section{Travaux connexes}\label{sec:relatedwork}

~\citet{he2016evaluation} présentent et comparent la précision et le temps de traitement de 4 méthodes de structuration sur 4 jeux de données issus d'applications différentes.
Leurs travaux mettent en valeur l'importance du temps de traitement et de la distribution des calculs pour 
tenir la charge dans de gros systèmes industriels. De la même façon, les travaux de synthèse de Logpai~\citep{zhu2019tools} évaluent la précision, la robustesse et l'efficacité de 13 méthodes sur 16 jeux de données.
La Table~\ref{tab_logparsing_methods} résume les caractéristiques des méthodes analysées adaptées au traitement en flux des \logs{}.
Les solutions les plus prometteuses en termes d'efficacité sont Drain, Spell et SHISO, Drain étant présenté comme la solution la plus performante selon ces mêmes critères.
L'article met en valeur la combinaison des résultats sur 16 jeux de données et l'impact fort du système d'information sur la précision des solutions.
\begin{table}[t]
  \small
 \begin{center}
   \begin{tabular}{l||c|c|c|c|c}
     {\it Log parser }  & Année & Technique & Efficacité & Couverture & Pré-traitement\\
   \hline
   SHISO & 2013 & Clustering & High & oui & non \\
   LenMa & 2016 & Clustering & Medium & oui & non  \\
   Spell & 2016 & Longest common subsequence & High & oui & non  \\
   Drain & 2017 & Parsing tree & High & oui & oui  \\
   \end{tabular}
\caption{ {\it Online log parsers} présentés dans \citet{zhu2019tools}.} \label{tab_logparsing_methods}
 \end{center}
\end{table}

Nous avons choisi de concentrer nos expériences sur les deux méthodes présentant la meilleure précision dans les études précédentes: Spell et Drain.
Dans le cadre de notre étude, nous avons ciblé et repris des jeux de données proches de l'environnement d'un fournisseur de Cloud (Tab.~\ref{tab_datasets}), 
les méthodes variant en fonction du système d'information dont provient chaque jeu de données."
OpenStack est une solution Cloud permettant un déploiement de plateformes IaaS; le volume du jeu de données reste cependant trop faible pour l'assimiler à un acteur industriel.
Nous avons retenu Android pour ses nombreux patrons différents et ses \logs{} de type système d'exploitation (allocation de ressources, gestion des processus et du réseau, etc.).
Pour finir, nous avons décidé de considérer HDFS, celui-ci étant largement utilisé dans la littérature malgré son aspect peu complexe et son faible nombre de patrons et de parties variables.
\begin{table}[t]
  \footnotesize
 \begin{center}
   \begin{tabular}{l||cccc}
     Jeu de données & Taille & \#Messages & \#Patrons & \#Mots uniques\\
   \hline
   OpenStack & 60,01 MB & 207,820& 51 & 942\\
   Android & 3,38 GB & 30,348,042 & 76,923 & 3599\\
   HDFS &  1,47 GB  & 11,175,629 & 30  & 1445\\
   \end{tabular}
\caption{Caractéristiques de certains jeux de données présentés dans~\citet{zhu2019tools}.} \label{tab_datasets}
 \end{center}
\end{table}
Les travaux présentés dans cette section forment une base solide pour comparer diverses méthodes existantes dans de nombreux environnements.
Toutefois, ils ne contiennent pas de données concrètes sur l'impact du paramétrage et du prétraitement sur la précision de l'analyse et le temps de calcul nécessaire pour traiter des jeux de données comme HDFS et Android.

%% file: setup.tex
\section{Évaluation}\label{sec:evaluation}

Nos travaux analysent l'impact du paramétrage et du prétraitement sur la précision et le temps de traitement des méthodes considérées.
Cette section est consacrée à la présentation et à l'analyse des résultats obtenus.

\paragraph{Contexte expérimental.}
Comme discuté dans la Sec.~\ref{sec:relatedwork}, nous avons retenu Spell~\citep{du2016spell} et Drain~\citep{he2017drain}, solutions les plus prometteuses en vue de leur application dans un environnement Cloud.
Deux métriques ont été choisies: la précision de structuration~\citep{zhu2019tools} définie comme le rapport entre le nombre de \logs{} dont le patron est correctement identifié et le nombre total de messages et le temps de calcul par {\it log} servant à représenter l'efficacité.
Trois jeux de données libres, disponibles sur la plateforme LogHub\footnote{https://github.com/logpai/loghub} ont été utilisés: OpenStack, Android et HDFS; des échantillons labellisés de 2 000 lignes ont servi pour le calcul de la précision, les jeux de données complets ont eux été utilisés pour l'évaluation du temps de calcul.
Nos expériences ont été réalisées sur une machine virtuelle Cloud de type CentOS Linux 7.8.2003 avec 62 GB de RAM.

%% file: parameters.tex
\subsection{Impact des paramètres sur la précision}

\begin{figure}[!tbh]
  \begin{subfigure}{\textwidth}
    \centering
  \begin{subfigure}{\textwidth}
    \centering
		\includegraphics[width=0.9\linewidth]{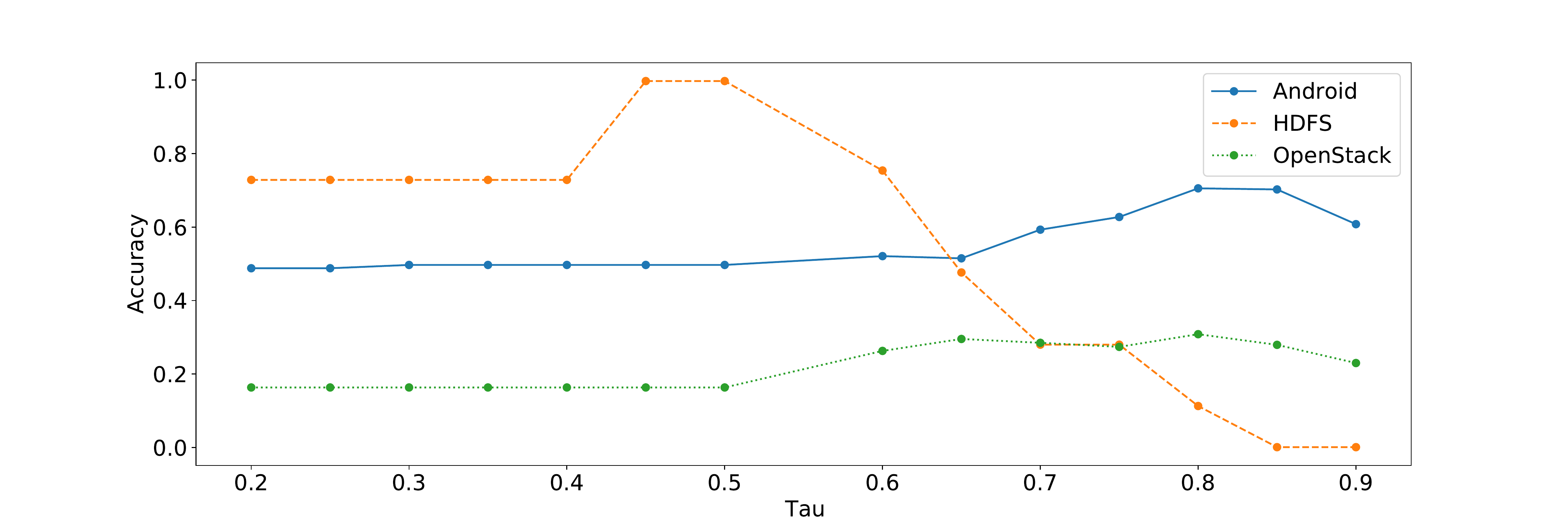}
		\caption{Spell}
	\end{subfigure}
  \begin{subfigure}{0.3\textwidth}
  		\centering
      \medskip
      Android
		\includegraphics[width=\linewidth]{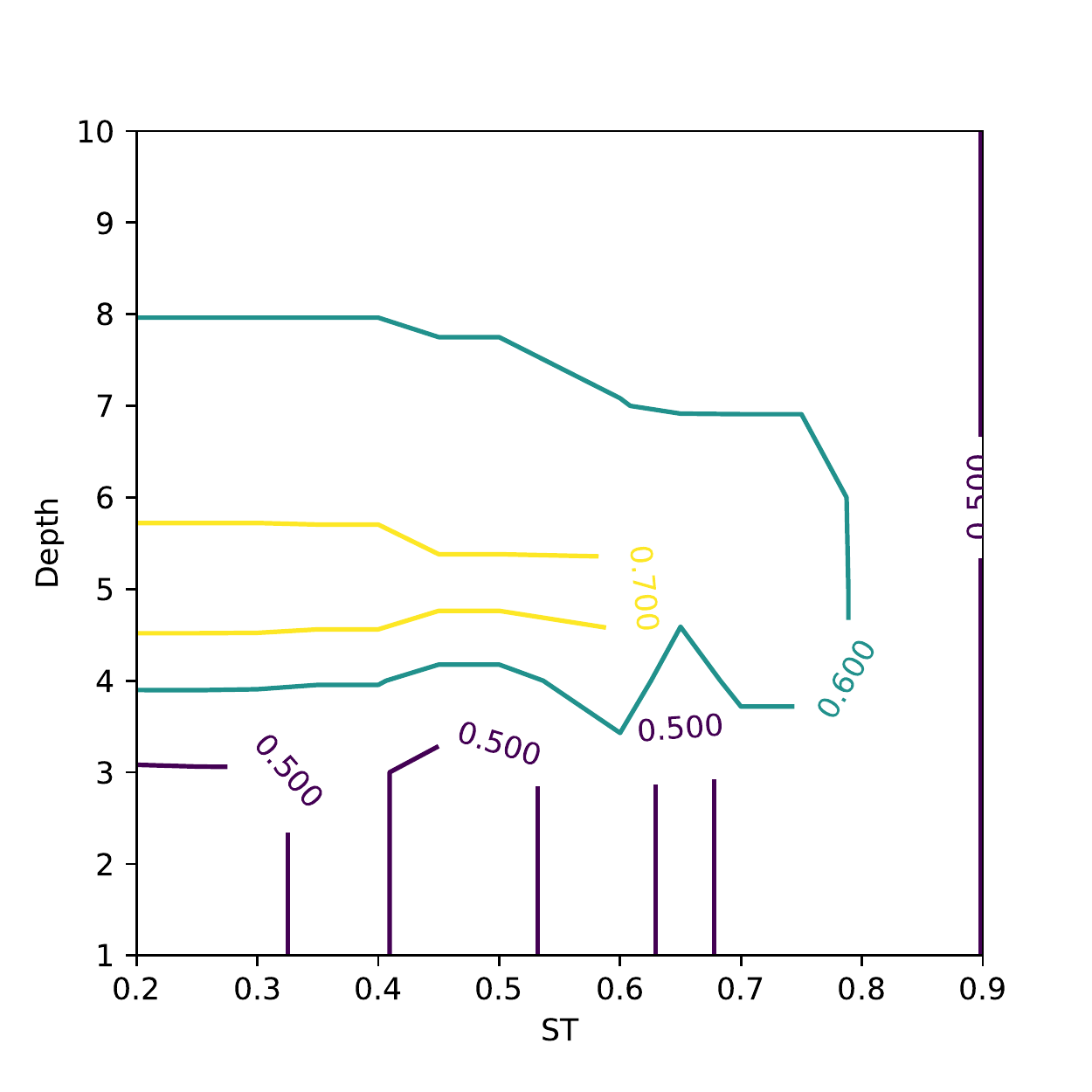}
	\end{subfigure}
  \begin{subfigure}{0.3\textwidth}
  		\centering
      \medskip
      HDFS
		\includegraphics[width=\linewidth]{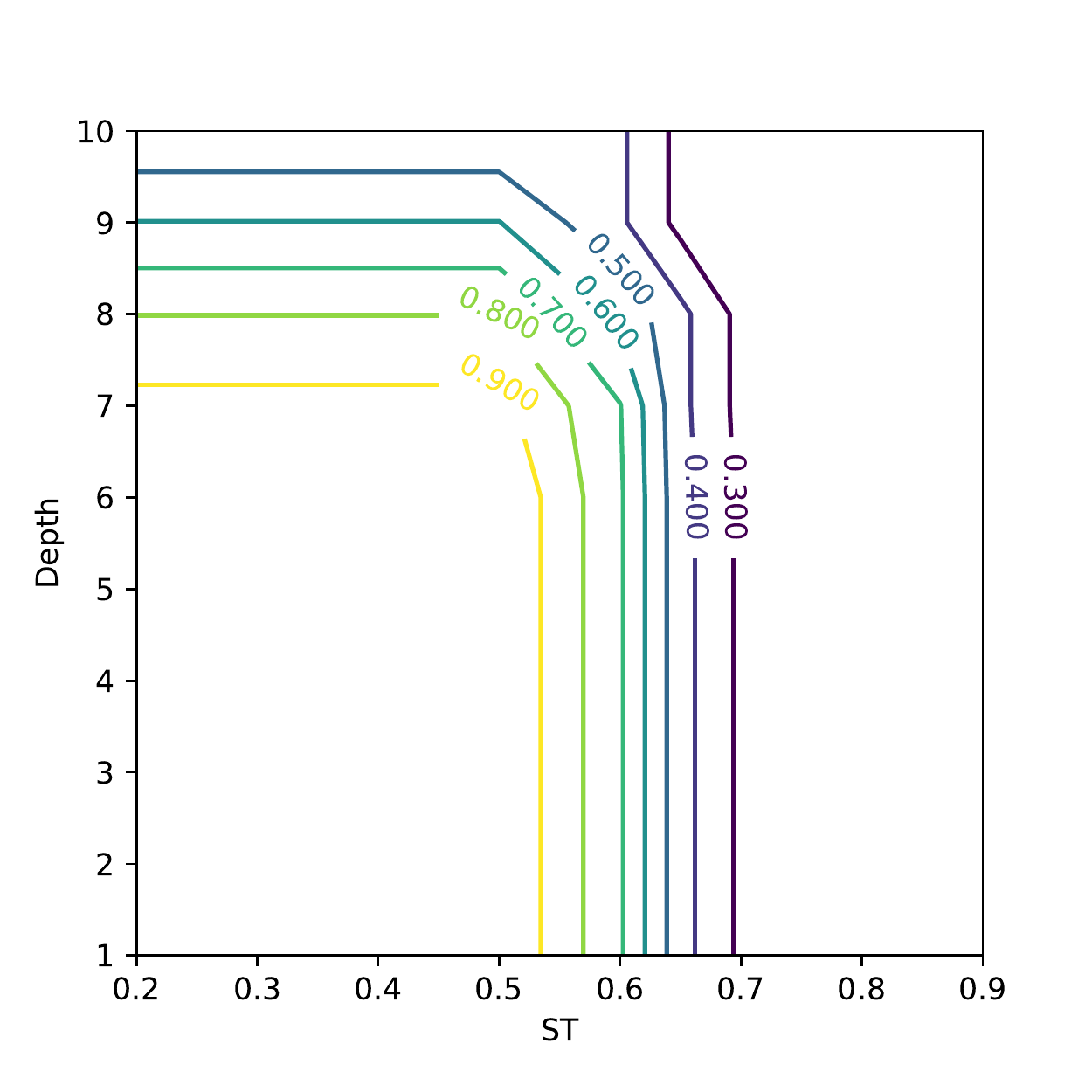}
	\end{subfigure}
	\begin{subfigure}{0.3\textwidth}
                \centering
	      \medskip
	            Openstack
        \includegraphics[width=\linewidth]{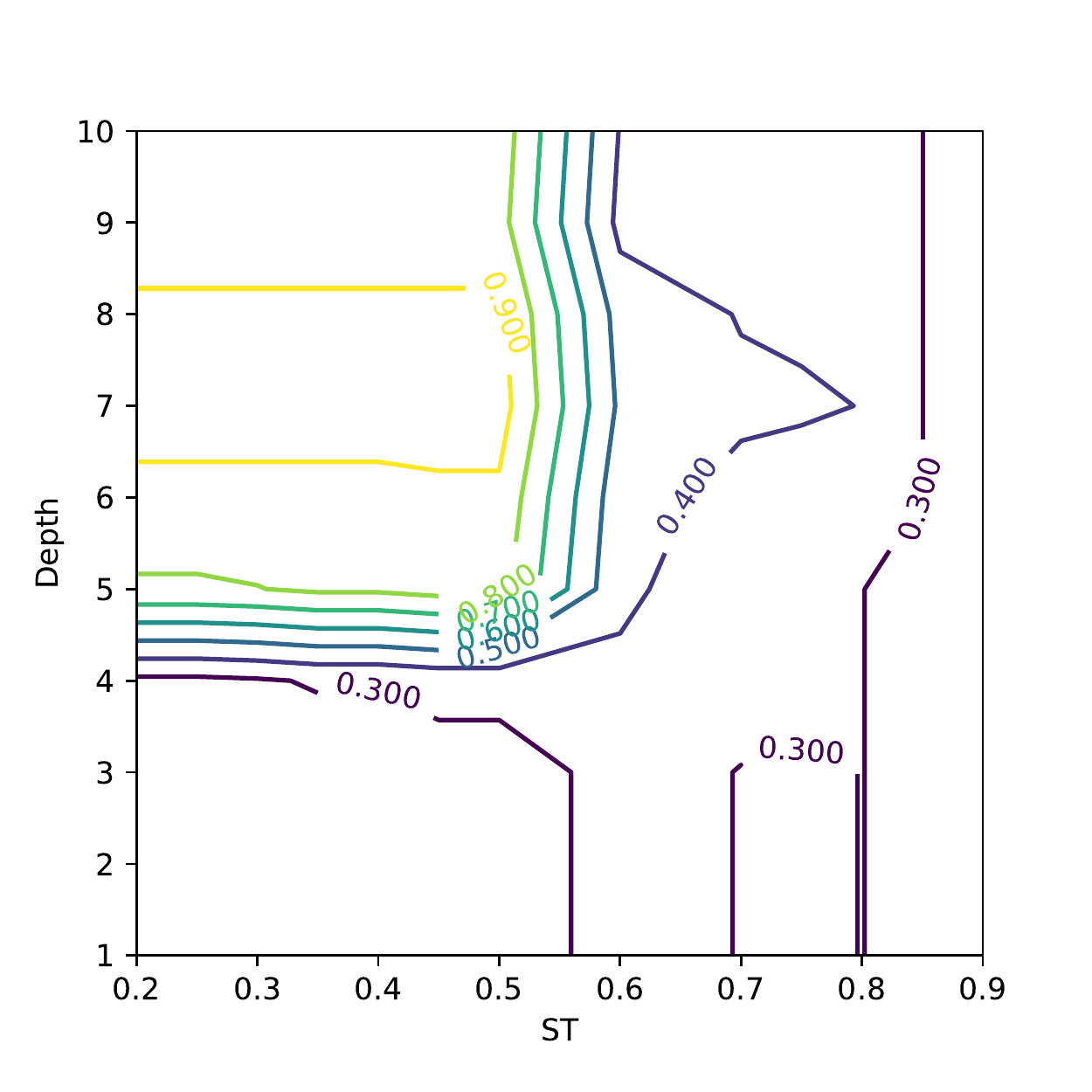}
        \end{subfigure}
  \caption{Drain}
  \label{fig:hyperparameters-spell}
\end{subfigure}

\caption{Précision en fonction des paramètres}
\label{fig:hyperparameters}
\end{figure}

Les paramètres sont des variables fixées lors de la configuration d'un système.
Dans cette section, nous présentons une étude de l'influence de ces paramètres sur la précision de Spell et Drain afin de savoir si des valeurs génériques sont envisageables.

Spell n'utilise qu'un seul paramètre servant de seuil pour déterminer si un \textit{log} appartient à un patron déjà connu, $\tau \in [0,1]$, qui est le rapport entre la plus grande séquence commune et la longueur du \textit{log}.
De son côté, Drain possède deux paramètres: une profondeur de l'arbre de recherche {\tt depth} et un seuil {\tt ST $\in [0,1]$} qui sert à déterminer si un \textit{log} appartient à un groupe existant.

La Figure~\ref{fig:hyperparameters} présente l'évolution de la précision en fonction des valeurs des paramètres.
La valeur de $\tau$ maximisant la précision de Spell diffère pour chaque jeu de données: 0,8 pour OpenStack, 0,85 pour Android et 0,5 pour HDFS.
Le jeu de données le plus touché par le choix de $\tau$ est HDFS : la moitié des valeurs de $\tau$ considérées donnent une précision inférieure à 0,7, pouvant même faire tomber la précision à 0.
Dans les deux autres jeux de données, le choix de $\tau$ peux faire gagner jusqu'à 15 \% de précision pour OpenStack et 20 \% pour Android.
Drain a un comportement inégal pour les différents jeux de données. Bien que des valeurs élevées de {\tt ST} (>0,7) ne donnent pas de bons résultats, la valeur de {\tt depth} est importante pour obtenir des résultats précis pour OpenStack et Android. 
De plus, dans le cas d'Android, la précision maximale obtenue est de 0,75 alors que 0,9 est atteint dans les deux autres jeux.
Pour les jeux de données concernés, Drain atteint la plus haute précision avec un {\tt ST} compris en 0,2 et 0,5 ; {\tt depth} en revanche est plus versatile, sa valeur optimale étant de 5 pour Android, 6 pour HDFS et 7 pour OpenStack.
On peut conclure que la précision des deux solutions est influencée par les choix de paramètres.

%% file: preprocessing.tex
\subsection{Impact du prétraitement sur la précision}

L'étape de prétraitement permet d'identifier des variables avant de commencer le processus de structuration d'un \textit{log}.
Nous avons sélectionné un ensemble d'expressions régulières (Tab.~\ref{tab:exprregulier}) identifiant des blocs
bien connus dans chaque jeu de données, tels que les chiffres, les adresses IP ou les mots commençant par "blk\_-" pour HDFS.

\begin{table}[t]
 \begin{center}
   \small
   \begin{tabular}{l||l}
     \multicolumn{1}{p{1.5cm}||}{\centering Jeu de \\ données} & \multicolumn{1}{|c}{\centering Expressions régulières } \\
    \hline
    \multirow{3}{*}{OpenStack} & {\tt \footnotesize ((\textbackslash{}d+\textbackslash{}.)\{3\}\textbackslash{}d+,?)+ } \\
     & {\tt \footnotesize /.+?\textbackslash{}s }\\
     & {\tt \footnotesize \textbackslash{}d+ }\\
   \hline
   \multirow{2}{*}{Android} & {\tt \footnotesize (/[\textbackslash{}w-]+)+', r'([\textbackslash{}w-]+\textbackslash{}.)\{2,\}[\textbackslash{}w-]+ } \\
    & {\tt \footnotesize \textbackslash{}b(\textbackslash{}-?\textbackslash{}+?\textbackslash{}d+)\textbackslash{}b|\textbackslash{}b0[Xx][a-fA-F\textbackslash{}d]+\textbackslash{}b|\textbackslash{}b[a-fA-F\textbackslash{}d]\{4,\}\textbackslash{}b } \\
   \hline
   \multirow{2}{*}{HDFS} & {\tt \footnotesize blk\_-?\textbackslash{}d+ } \\
    & {\tt \footnotesize (\textbackslash{}d+\textbackslash{}.)\{3\}\textbackslash{}d+(:\textbackslash{}d+)? } \\
   \end{tabular}
\caption{Expressions régulières utilisées dans le prétraitement} \label{tab:exprregulier}
 \end{center}
\end{table}

\begin{table}[t]
 \begin{center}
   \footnotesize
   \begin{tabular}{l||ll|ll|ll|ll}
     Jeu de & \multicolumn{2}{|c}{\centering Spell (sans)} & \multicolumn{2}{|c}{\centering Spell (avec)} & \multicolumn{2}{|c}{\centering Drain (sans)} & \multicolumn{2}{|c}{\centering Drain (avec)}  \\
     données   & \multicolumn{1}{|c}{prec.} & nb. patrons & \multicolumn{1}{|c}{prec.} & nb. patrons & \multicolumn{1}{|c}{prec.} & nb. patrons & \multicolumn{1}{|c}{prec.} & nb. patrons \\
   \hline
   Android   & 0.60&425 & 0.91(x1.5)&180(x0.42) & 0.67&217& 0.91(x1.4)&171(x0.79) \\
   HDFS      & 0.28&684 & 1.00(x3.6)&14(x0.02) & 1.00&17 & 1.00(x1)&16(x0.94) \\
   O.Stack & 0.23&692 & 0.77(x3.3)&451(x0.65) & 0.84&75  & 0.73(x0.8)&299(x3.99) \\
   \end{tabular}
\caption{ Précision en fonction du prétraitement} \label{tab:pretraitement}
 \end{center}
\end{table}

La Table~\ref{tab:pretraitement} présente les résultats obtenus sur les échantillons de 2 000 lignes.
Chaque méthode a tourné avec et sans prétraitement,
nous avons calculé la précision, mais également le nombre de patrons afin de détecter une éventuelle surclassification des \logs{}.
Les paramètres utilisés sont ceux pour lesquels chaque méthode présente la plus haute précision avec prétraitement.

Spell arrive à multiplier par 1,5 sa précision avec l'utilisation du prétraitement, et même par  3,6 dans le cas de HDFS. Même constat pour les patrons : Spell bénéficie d'une influence positive en se rapprochant du nombre réel.
Pour Drain, les résultats diffèrent selon le jeu de données, tandis que pour Android, nous notons une amélioration une amélioration similaire à celle de Spell avec le prétraitement. Pour OpenStack, la précision et le nombre de patrons empirent avec les mêmes expressions régulières grâce auxquelles on a constaté des améliorations dans le cas de Spell.
Cette baisse de précision est liée à des mots de la forme 54b44eb-2d1a-4aa2-ba6b-074d35f8f12c présents dans 3 patrons. Sans prétraitement, celui-ci altère un niveau de l'arbre de partitionnement. En revanche, avec prétraitement ce mot est découpé en plusieurs parties: *b*eb-*d*a-*aa*-ba*b-*d*f*f*c et altère plusieurs niveaux de l'arbre
, faisant ainsi tomber à 0 la précision de Drain pour tout \textit{log} contenant ce type de mot. Spell n'est pas affecté car sa précision sur les patrons contenant ce mot est déjà nulle.

Le prétraitement a donc un effet, mais pas toujours positif sur la précision. Cette étape est de plus coûteuse car exigeant de déterminer à l'avance les expressions régulières.

%% file: timeprocessing.tex
\subsection{Impact sur le temps de calcul}

Notre dernière évaluation porte sur le temps du traitement: celui-ci doit rester le plus bas possible pour espérer des résultats en temps réel.
Nous avons dans un premier temps essayé de faire tourner Spell et Drain sans prétraitement mais le temps d'exécution de Spell était trop long. En effet, au bout de trois heures, Spell n'avait pas terminé les calculs et ce pour les trois jeux de données contre 
quelques dizaines de minutes avec prétraitement. Les valeurs présentées dans cette section concernent donc les résultats avec étape de 
prétraitement.

\begin{figure}[t]
  \centering
  \begin{subfigure}{0.29\textwidth}
    \centering
  \medskip
  Android
		\includegraphics[width=\linewidth]{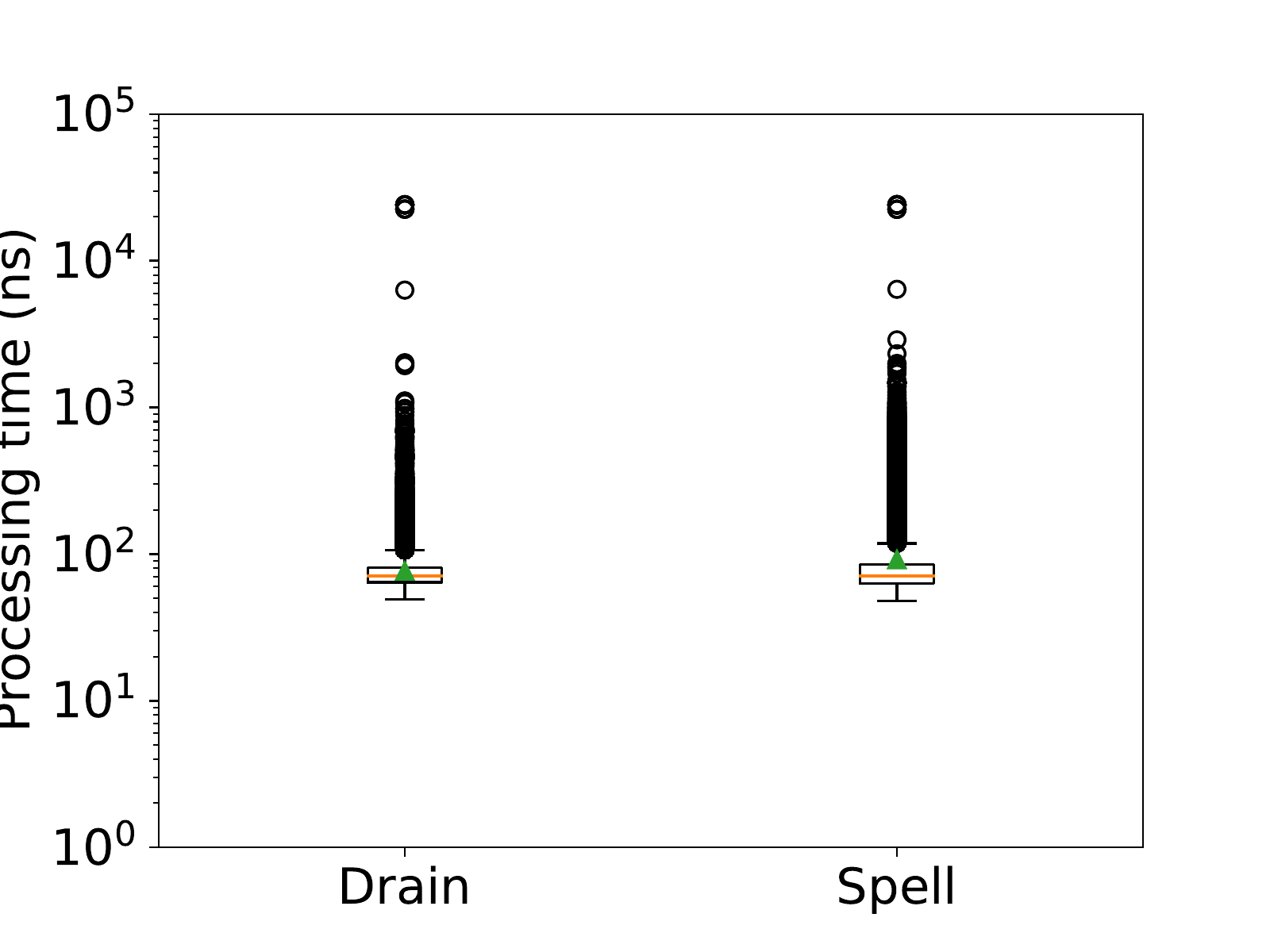}
	\end{subfigure}
  \begin{subfigure}{0.29\textwidth}
    \centering
  \medskip
  HDFS
		\includegraphics[width=\linewidth]{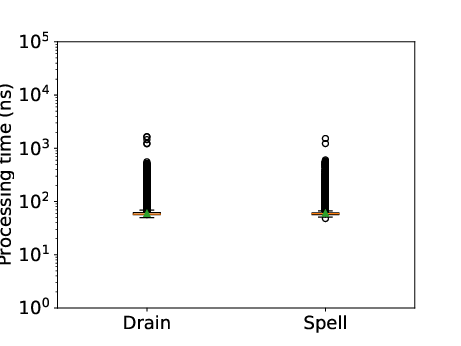}
	\end{subfigure}
  \begin{subfigure}{0.29\textwidth}
    \centering
  \medskip
  OpenStack
                \includegraphics[width=\linewidth]{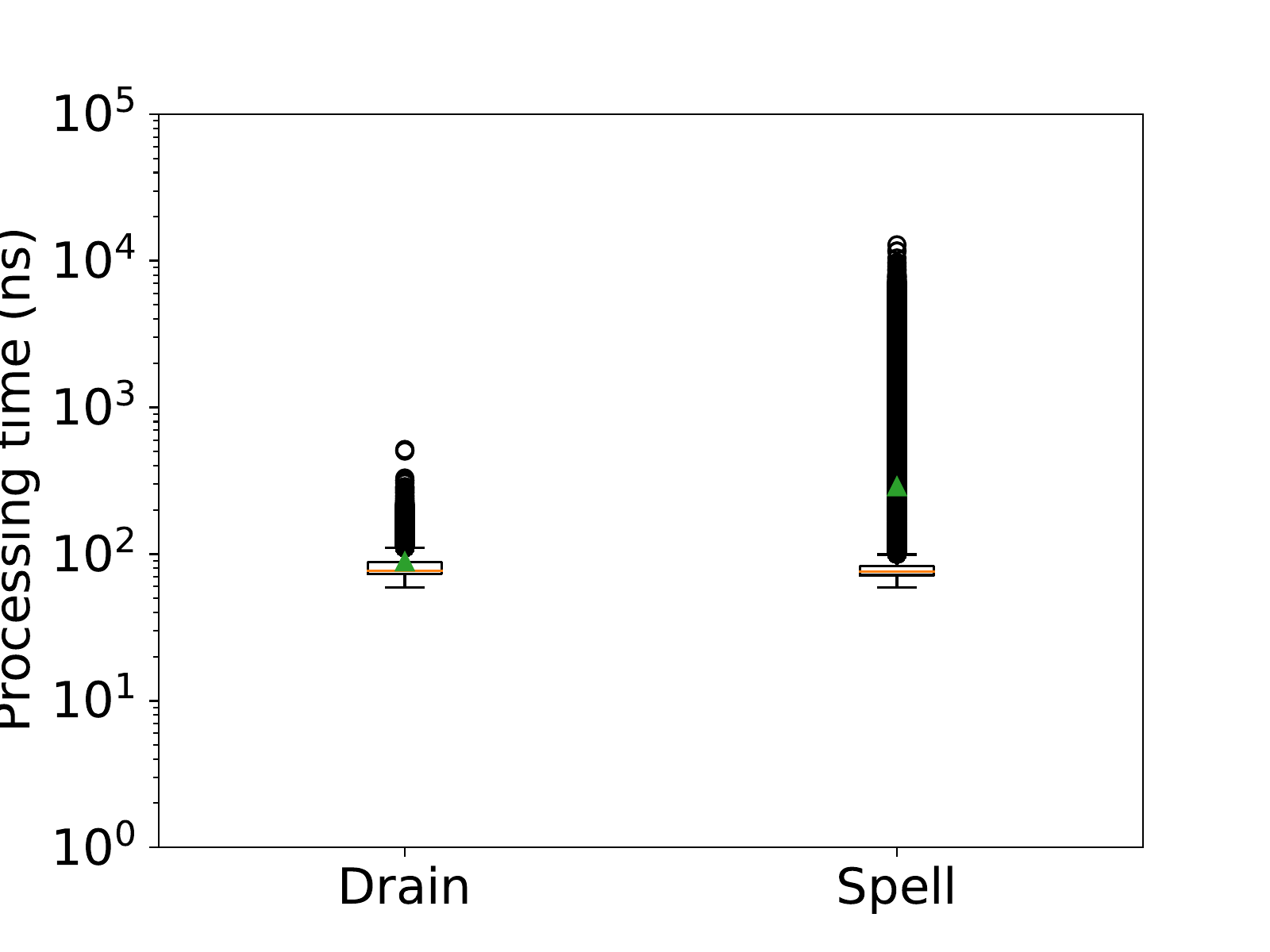}
        \end{subfigure}
\caption{Temps de traitement total en fonction du nombre de logs déjà structurés}
\label{fig:time}
\end{figure}

Les résultats présentés dans la Figure~\ref{fig:time} montrent la distribution du temps de traitement sur les différents jeux de données.
Le temps moyen de traitement de Drain reste dans un ordre de 10**2 ns sur chaque jeu de données et est toujours inférieur à celui de Spell. 
Il est également important de noter qu'il existe un nombre conséquent de \logs pour lesquels le temps de traitement peut être jusqu'à 500 fois supérieur à la moyenne.
Il est important de noter ces temps divergents car ils sont susceptibles de créer des décalages par rapport au temps réel dans un environnement de production.

%% file: conclusion.tex
\section{Conclusion et travaux futurs}\label{sec:conclusion}

L'analyse des fichiers de \logs{} est primordiale pour la maintenance des systèmes informatiques et l'automatisation de celle-ci passe par une étape de structuration visant à exploiter l'information contenue dans les messages.
Dans cet article, nous nous sommes intéressés à la robustesse
et l'efficience de deux méthodes issues de la littérature récente.
Ces deux aspects sont cruciaux dans un environnement Cloud où l'interaction humaine doit être minimale pour gérer un volume et une variabilité conséquente des \logs{}.
Nos résultats montrent que Spell et Drain sont deux approches prometteuses. Cependant, toutes deux sont sensibles au paramétrage, pouvant faire tomber la précision à 0 dans certains cas.
Les deux méthodes sont également influencées par l'utilisation d'une étape de traitement préliminaire : Spell en bénéficie fortement, là où Drain se comporte de manière inégale, avec l'existence d'un impact négatif sur le jeu de données OpenStack.
Finalement, nous mettons en valeur l'existence d'un écart conséquent par rapport au temps de traitement moyen pour de nombreux \logs{} sur tous les jeux de données. Cet écart se manifeste pour Spell et Drain,
et pourrait, dans un environnement aux volumétries exigeantes, causer des retards et donc altérer un traitement en temps quasiréel.

Suite à cette étude et afin de satisfaire les contraintes d'une solution de structuration des \logs{} dans un environment Cloud, nous allons explorer des solutions autonomes et indépendantes du paramétrage et du prétraitement, à même de conserver une précision élevée et un temps de traitement linéaire.